\documentclass[a4paper]{article}

\usepackage{INTERSPEECH2018}
\usepackage{amsmath}
\usepackage{breqn}
\usepackage{color}
\usepackage{multirow}
\usepackage{url}
\newcommand{\R}{\mathbb{R}}

\title{Detecting Alzheimer's Disease Using Gated Convolutional Neural Network from Audio Data}
\name{Tifani Warnita, Nakamasa Inoue, Koichi Shinoda}
\address{
  Tokyo Institute of Technology, Tokyo, Japan}
\email{\{tifani,inoue\}@ks.cs.titech.ac.jp, shinoda@c.titech.ac.jp}

\begin{document}

\maketitle
\begin{abstract}


We propose an automatic detection method of Alzheimer's diseases using a gated convolutional neural network (GCNN) from speech data. This GCNN can be trained with a relatively small amount of data and can capture the temporal information in audio paralinguistic features. Since it does not utilize any linguistic features, it can be easily applied to any languages. We evaluated our method using Pitt Corpus. The proposed method achieved the accuracy of 73.6\%, which is better than the conventional sequential minimal optimization (SMO) by 7.6 points.

\end{abstract}
\noindent\textbf{Index Terms}: Alzheimer's disease, computational paralinguistics, convolutional neural network, gating mechanism

\section{Introduction}

Alzheimer's Disease (AD) is the most common cause of dementia \cite{alz2017}, a neurodegenerative disease strongly related to the reduced functionality or even death of neurons in the central nervous system \cite{yacoubian2017neurodegenerative}. As the result of ageing society, we face an increasing number of people being affected by AD \cite{fraser2016linguistic} which is estimated to be doubled every 20 years \cite{ferri2005global}. The most noticeable symptom of AD patients is the memory loss \cite{burnsb158} such as in recalling experiences, which results in poor narrative memory \cite{prud2011extraction}. They also often become apathy and easy to get depressed \cite{alz2017}.

At present, there is no clear protocol on how to detect AD not only in an effective but also accurate way \cite{alz2017}. The most common approach is to monitor the patients, to carefully examine the medical history of the patients, to conduct some cognitive tests (i.e. a picture description task, a naming task), mental status, and mood test, and to take their brain images. The careful diagnosing process can take in days or even in weeks which might be very cumbersome. Early prediction of AD actually can help its patients to preserve their cognitive functions \cite{martono2000buku}. Some of their causes are treatable and the patients are sometimes fully recovered \cite{tripathi2009reversible}. Automatic detection of AD in its early phase has been strongly demanded.

To date, there have been a lot of studies for automatically detecting AD patients. Most of them used linguistic information \cite{fraser2016linguistic,wankerl2017n}, which is difficult to be applied to different languages, especially to low-resource languages. In this study, we propose a non-linguistic approach for detecting AD using acoustic features from speech data. Inspired by numerous successes with deep learning for paralinguistic tasks such as for emotion recognition \cite{huang2014speech, keren2016convolutional}, we employ convolutional neural networks with a gating mechanism.

\section{Related Works}

Linguistic information had been widely utilized in order to automatically detect AD patients \cite{zimmerer2016formulaic,orimaye2017predicting,wankerl2017n}. For instance, Wankerl et al. (2017) \cite{wankerl2017n} employed a pure linguistic approach based on $n$-gram on Pitt Corpus \cite{becker1994natural}, in which subjects undergo a picture description task. From the study, they found that patients often not only uttered incomplete phrases but also interrupted others. These degraded the intelligibility of their speech \cite{khodabakhsh2015evaluation}.

Some studies utilized both of linguistic features and acoustic features to detect people having AD \cite{khodabakhsh2015evaluation,fraser2016linguistic,sadeghian2017speech}. Fraser et al. (2016) \cite {fraser2016linguistic} utilized multilinear logistic regression and selected 35 top-ranked features out of 370 features by using Pitt Corpus \cite{becker1994natural}. Khodabakhsh et al. (2015) \cite{khodabakhsh2015evaluation} used the recordings of conversational speech of 32 AD patients and 51 control subjects for this combination. Weiner et al. (2016) \cite{weiner2016speech} employed only acoustic information from German conversational recordings. However, the study was relied on the transcription for calculating some features such as silence-to-word ratio and word rate.

In contrast with these studies, we focused on non-linguistic approach where only speech audio features of the subjects are utilized. It can be easily applied to other languages and expected to be more robust against environmental noise and channel differences. As features, we used the paralinguistic features often used in emotion recognition task since the previous studies showed that AD people tend to show emotional prosodic impairment \cite{tosto2011prosodic}. Numerous approaches for detecting emotions emerged from people have been carried out. Schuller et al. (2009) \cite{schuller2009interspeech} provided the baseline for emotion recognition in the INTERSPEECH 2009 Emotion Challenge, which uses sequential minimal optimization (SMO). Huang et al. (2014) \cite{huang2014speech} employed deep learning based approach in which convolutional layers were employed. Keren and Schuller (2016) \cite{keren2016convolutional} used not only convolutional but also recurrent layers.

\section{Database}

In this study, we used the picture description task session data of Pitt Corpus \cite{becker1994natural}, which is a part of DementiaBank, a multimedia database for studying people having dementia. In the picture description task, patients are asked to describe what happens in a picture, Cookie Theft Picture of the Boston Diagnostic Aphasia Examination \cite{kaplan1983boston}. Pitt Corpus consists of speech data and their transcription from 244 control (healthy) subjects as well as 309 patients having dementia such as mild cognitive impairment (MCI), vascular dementia, and AD. In this study, we used only the data from AD patients and from patients suspected of having AD (probable AD). It should be noted that, even though we used the same database as in \cite{fraser2016linguistic,wankerl2017n}, the number of the subjects in our study was slightly different from theirs from the following three reasons. First, the size of the database has increased over time. Second, we excluded speech data with overlapped sounds from the other interview sessions. Third, we only used those data with both audio data and transcription information.  Consequently, the number of sessions is 488 (255 AD, 233 control) from 267 subjects (169 AD, 98 control). Similar to \cite {fraser2016linguistic}, due to a limited amount of data, we employed a 10-fold cross validation scheme for our evaluation. We designed the ten subsets so that no subject appeared in more than one subset in order to avoid any speaker dependencies. 

In this study, we performed three preprocessing stages. First, we normalize each audio signal using the average value of decibels relative to full scale (dBFS) in the data. Then, we segmented the audio data of each subject into utterances according to the transcription information. We obtained the total of 6267 utterances (3276 AD, 2991 control). Finally, we added 10ms at the beginning and at the end of each utterance segment with 15ms fade-in and fade-out.

\section{Features}
\label{sec:features}

In this study we used openSMILE \cite{eyben2013recent} to extract acoustic features from Pitt Corpus. OpenSMILE consists of several different configurations of acoustic feature extraction. It has been mainly used for emotion recognition but we believe it is also effective for our purpose. While AD patients can easily get depressed, anxious, or even upset \cite{alz2017}, they find it difficult to express their emotions in prosodies such as tempo alteration and powerful intonation \cite{tosto2011prosodic}. Based on this finding, we use the following feature sets.
 
\begin{enumerate}
\item{INTERSPEECH 2009 Emotion Challenge Features (IS09) \cite{schuller2009interspeech}}

In this feature set, there are 16 types of low-level descriptors (LLD) extracted from the frame level. The delta coefficients were also calculated hence producing the total of 32 LLD. In order to get the utterance-level features from LLD, 12 functionals are introduced (e.g. the values of minimum and maximum, mean, and range) are applied to each LLD. As a result, 384 features are extracted from one utterance.

\item{INTERSPEECH 2010 Paralinguistic Challenge Features (IS10) \cite{schuller2010interspeech}}

The additional LLD to IS09 are PCM loudness, eight log Mel frequency band (0-7), eight line spectral pairs (LSP) frequency (0-7), F0 envelope, voicing probability, jitter local, jitter DPP, and shimmer local. In addition, more MFCC features are extracted (0-14 compared to 1-12). Finally, we get 76 LLD for one frame and the total of 1582 features for one utterance.

\item{INTERSPEECH 2011 Speaker State Challenge Features (IS11) \cite{schuller2011interspeech}}

Compared to the previous feature sets, IS11 provides the derived loudness measure and the employment of Relative Spectral Analysis (RASTA)-style filtered auditory spectra resulting in 118 LLD for the frame-level features. The total number of features in one utterance is 4368.

\item{INTERSPEECH 2012 Speaker Trait Challenge Features (IS12) \cite{schuller2012interspeech}}

In this feature set, some of the LLD added compare to IS11 includes harmonic-to-noise ratio (HNR), spectral harmonicity, and psychoacoustic spectral sharpness resulting in 120 LLD for the frame-level features. After being applied by functionals, total number of features in one utterance is 6125 features.

\end{enumerate}

\section{Method}
\begin{figure}[t]
  \centering
  \includegraphics[width=80mm,scale=0.5]{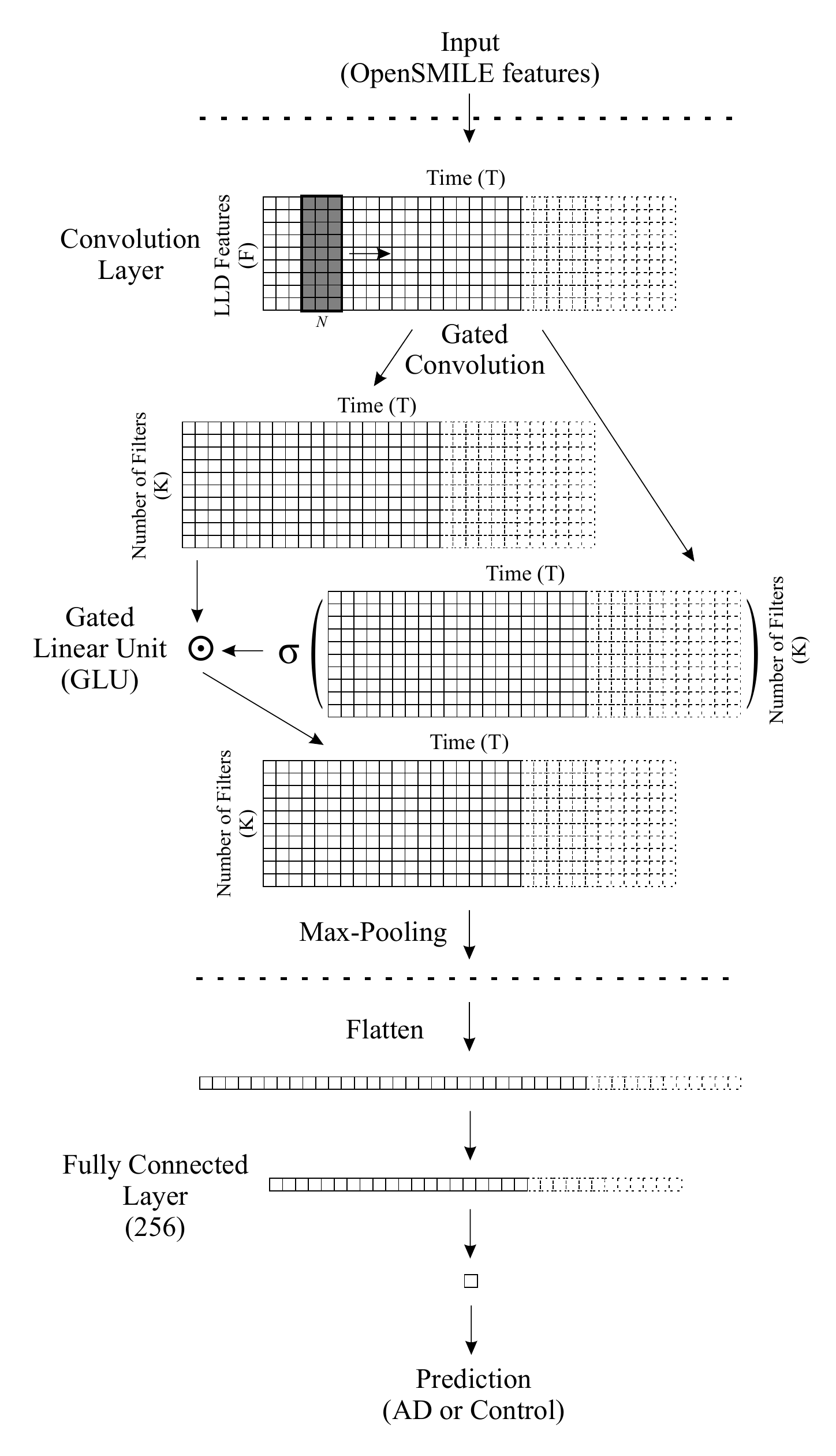}
  \vspace{-1em}
  \caption{GCNN with depth = 1. The kernel window in the convolutional layer is colored in gray.}
  \label{fig:framework_ov}
  \vspace{-1.5em}
\end{figure}

\subsection{Convolutional Neural Network}
\label{sec:cnn}
In this study, assuming that temporal features are well represented in the frame-level features, we employed a Convolutional Neural Network (CNN) \cite{lecun1989backpropagation} as a classifier. A CNN was inspired by the structure of animal visual cortex for perceiving lights \cite{gu2017recent} and has yielded supreme results in an abundance of tasks in the past few years \cite{goodfellow2016deep}. Furthermore, it needs relatively small amount of training data compared to the other networks since it has a much smaller number of connection weights \cite{bengio2009learning}.

The convolution operation over the input aims to extract the temporal information by sliding through the use of a kernel (filter). In our study, we feed utterance segment features \(X \in \R ^ {F \times T}\) into the CNN, where \(F\) and \(T\) represent the dimension of a feature vector of LLD and the number of time frames, respectively. The size of the sliding window is \(\R ^ {F \times N}\) where \(N\) is the window length of the kernel in time axis. The convolution operation between the kernel and the input will produce a scalar output. Output from the convolution layer at time \(i\), \(i = n+1, ..., T\) is then defined as,

\begin{equation}
\label{eq:cnn1}
y_{i} = g\left(\sum\limits_{f = 1}^{F} \sum\limits_{n = 1}^{N} w_{f,n}x_{f,i-n} + b\right),
\end{equation}

\noindent
where \(b\) and \(g\left(\cdot\right)\) denotes the bias and activation function respectively. \(x_{c,d}\) is the \((c,d)\) element of the input \(X\), and \(w_{c,d}\) is the \((c,d)\) element of the weight matrix \(W\). Both \(W\) and \(b\) are the learnable parameters of the network that we train. When the kernel is sliding through the input matrix over the time dimension, we multiply the overlapping element of the two matrices as in Eq. \ref{eq:cnn1}. We employed a rectified linear unit (ReLU) \cite{hahnloser2000digital} as the activation function \(g\). This network is also referred as Time-Delay Neural Network (TDNN) \cite{waibel1990phoneme}.

Since the input dimension for CNN should be fixed, we set the segment length as that of the longest utterance segment in the dataset. After that, we applied zero padding for the rest of the utterance segment. We use the number of kernels \(K = 64\) and the window length of \(N = 3\). Accordingly, the calculation of every patch of the input will produce the complete output \(Y \in \R ^ {M \times K}\) where \(M\) is the number of time frames, \(T - N + 1\).

We added batch normalization before each activation function \cite{ioffe2015batch}. We also used random weight initialization for each convolution layer. After each convolution layer, we put a max-pooling layer \cite{zhou1988computation}. 

The output of the last convolution layer is flattened into one feature vector. For example, the flattened output \(Y \in \R ^ {M \times N}\) will produce a vector \(Z \in \R ^ O\) where \(O = M \times N\). This vector became the input to a fully-connected layer with activation function ReLU consisting 256 hidden neurons. We also employed batch normalization and initialized the weight matrix as random uniform. Furthermore, we also applied dropout with the value of 0.5 before the output layer. The output layer consists one hidden neuron with a sigmoid function. We used binary cross-entropy as the loss function and Adam \cite{kingma2014adam} as the optimizer. We trained the network with the maximum number of epochs 200 based on the pair of LLD features of each utterance segment and its corresponding binary label. 

\subsection{Gating Mechanism}

In addition to the standard deep convolutional neural network, we introduce a gating mechanism after each convolution. The resulting network is called Gated Convolution Neural Network (GCNN). A gate represents an information controller which manages the information that flows into the succeeding layer. The gate function can prevent the gradient from being vanished during backpropagation \cite{dauphin2016language}. Recently, it has been often used in convolutional neural networks such as for conditional image modeling \cite{oord2016conditional}, language modeling \cite{dauphin2016language}, and speech synthesis \cite{oord2016wavenet}. Previous study \cite{dauphin2016language} showed that gated linear unit (GLU) outperformed gated tanh unit (GTU) which is used in \cite{oord2016conditional}. Therefore, we employed GLU in our study.

Similar as in Section \ref{sec:cnn}, we give the input of extracted LLD features of each utterance segment \(X \in \R ^ {T \times F}\). We used a sigmoid function as the activation function \(g\), which is multiplied by a linear gate. Eq. \ref{eq:cnn1} is modified as:

\begin{dmath}
\label{eq:gcnn}
y_i = \left(\sum\limits_{f = 1}^{F} \sum\limits_{n = 1}^{N} v_{f,n}x_{f,i-n} + e\right)
\cdot g\left(\sum\limits_{f = 1}^{F} \sum\limits_{n = 1}^{N} w_{f,n}x_{f,i-n} + b\right),
\end{dmath}

\noindent
where \(v_{c,d}\) represent the element of \(V\) at position \((c,d)\). The \(V \in \R ^ {K \times F}\) and \(e \in \R\) are the kernel weight matrix and bias for the linear gate. For GCNN, we used \(N = 2\) and applied the same padding as in \cite{oord2016wavenet}.

After that, we halved the output length in a max-pooling layer \cite{zhou1988computation}. Figure \ref{fig:framework_ov} shows the visualization of our GCNN with the depth \(= 1\) where the \(depth\) represents the number of gated convolution layers. In the figure, one gated convolution layer lies between the dotted horizontal line which followed by the max-pooling layer. Deeper network consists of more gated convolution layers. We also defined the same layers after the max-pooling layer and used the same number of epochs as in Section \ref{sec:cnn}.

\subsection{Framework Overview}
While we need to classify each subject based on his/her whole data, we performed the classification for each utterance instead. This is based on our assumption that the information about a patient having AD or not can still be obtained from a smaller but appropriate length of a segment. After the utterance-based classification is performed, we make the final verdict for each subject based on the proportion of each class; we classified a subject into AD if he/she has AD percentage above the half. The utterance-level subject classification is expected to give more detailed information about the symptoms while most of the previous studies conducted the subject-level classification \cite{khodabakhsh2015evaluation,fraser2016linguistic,zimmerer2016formulaic,orimaye2017predicting,wankerl2017n,sadeghian2017speech}.

\section{Experimental Results}

\begin{figure}[t]
  \centering
  \includegraphics[width=75mm]{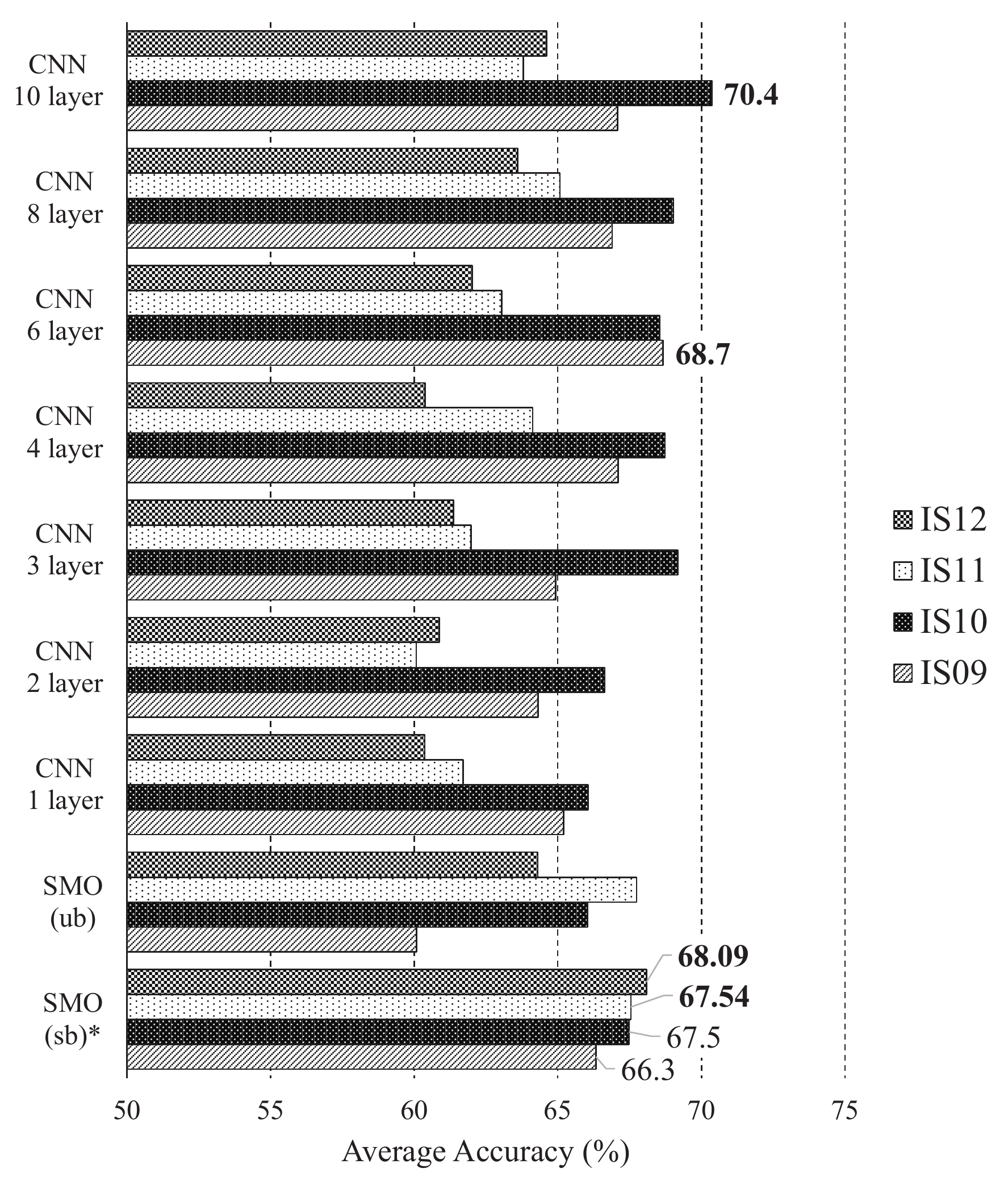}
  \caption{Comparison of IS09, IS10, IS11, and IS12 feature sets using standard CNN with sb and ub denote the subject-level and utterance-level based classification respectively.}
  \label{fig:std_cnn_comp}
\end{figure}

\begin{figure}[t]
  \centering
  \includegraphics[width=\linewidth]{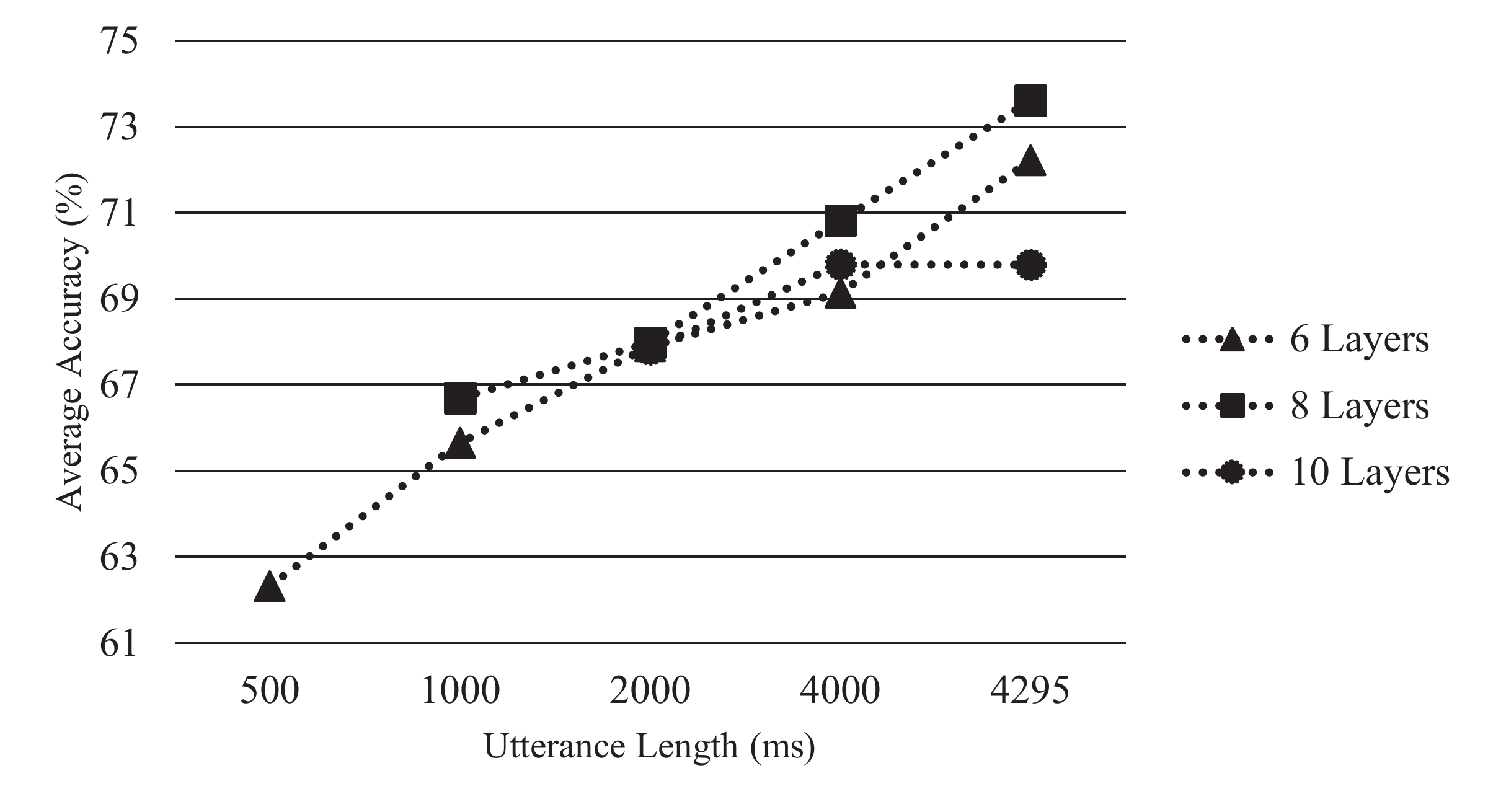}
  \caption{GCNN with different utterance length.}
  \label{fig:uttr_length_stat}
\end{figure}

\begin{table}[t]
\centering
\caption{Comparison of the employment of the standard CNN and GCNN by using IS10 feature set represented in the average accuracy (\%) from 10-fold cross validation.}
\label{table:all_comparison}
\begin{tabular}{@{}ccccc@{}}
\toprule
\textbf{Depth} & \multicolumn{2}{c}{\textbf{Standard CNN}} & \multicolumn{2}{c}{\textbf{Linear GCNN}} \\ 
\textbf{}      & \textbf{Utterance}  & \textbf{Subject}  & \textbf{Utterance}   & \textbf{Subject}  \\ \midrule
1              & 64.2              & 66.0            & 62.2               & 66.2            \\
2              & 64.2              & 66.6            & 62.6               & 68.7            \\
3              & 64.2              & 69.2            & 61.9               & 66.4            \\
4              & 64.9              & 68.7            & 63.3               & 68.9            \\
6              & 65.5              & 68.6            & 65.1               & 72.2            \\
8              & 66.1              & 69.0            & 66.3               & \textbf{73.6}   \\
10             & 65.2              & 70.4            & 65.2               & 69.8            \\ \bottomrule
\end{tabular}
\end{table}

We used the average accuracy from the 10-fold cross validation scheme on Pitt Corpus \cite{becker1994natural} (see Section 3 for details) for evaluating our method. Our first experiment was the employment of the standard deep CNN without gates on the four feature sets, IS09, IS10, IS11, and IS12. Figure \ref{fig:std_cnn_comp} shows the classification result of AD/non AD with different numbers of hidden layers (1, 2, 3, 4, 6, 8, 10). The performance result given is from the subject-level classification after majority voting from the utterance-level classification. Since there is no established baseline for this database and the number of instances used are different from one experiment to another experiment, we give the result of the four feature sets with baseline methods, which is the subject-level classification using sequential minimal optimization (SMO) \cite{schuller2009interspeech,schuller2010interspeech}. In Figure \ref{fig:std_cnn_comp}, they are marked by a star symbol (*). 

From Figure \ref{fig:std_cnn_comp}, the best result is obtained when we use 10-layer CNN with IS10 feature set which is better than SMO by 2.4 points. Furthermore, we can see that the use of CNN with both IS11 and IS12 could not yield better result compared to using SMO while it can improve the performance over using IS09 and IS10. We can also see that the IS10 feature set outperformed the rest of the feature sets when we use CNN.

When we compare between IS09 and IS10, we can see that IS10 covers more features in the paralinguistic aspect of speech as it was used as the age-gender and level-of-interest classification. The performance of the feature sets IS09 was worse than that of IS10 especially when the subjects did not have any specific emotion (neutral). Some emotions appeared at the beginning before the subject begins to describe the image, in the middle when the subject begins to confuse with the things he/she wants to describe (laughs), and at the end after completing the task.

Noticeable differences between IS09 and IS10 include the use of voicing probability in which might more represent the sound-silence pattern in the subjects. Further, jitter and shimmer in IS10 might give more representation of the hesitation rate in the subjects. Those LLD are also appear in both IS11 and IS12. However, the higher dimension of the two feature sets might be too big for the input of the CNN.

Next, we investigated the effectiveness of the gating mechanism. The experiment was carried out by using IS10 feature set. The comparison of the standard CNN and the gating mechanism is shown in Table \ref{table:all_comparison}. From the table, we can see that the employment of linear gate improved the average accuracy from the 10-fold cross validation scheme into 73.6\%. We can see that the gating mechanism yields better result.

Lastly, we investigated the importance of utterance length information in the classification performance. We tried a set of different segment length $L$, which are 500ms, 1000ms, 2000ms, 4000ms, and also 4295ms which is the segment length of the longest utterance in the dataset. In this case, we segmented each subject data into segments with a predetermined length $L$ without taking into consideration the oracle utterance length. Accordingly, the zero padding is added only for the last utterance segment of the subject if its length is less than $L$. The experiment was carried out by using the best scheme from the previous experiment, which is GCNN. We only tried to use the gated CNN with the number of layers as 6, 8, and 10.

As depicted in Figure \ref{fig:uttr_length_stat}, we can see that shorter segment length yields worse results. However, we can still get close results using the utterance length of 4000ms (69.1\%, 70.8\%, and 69.8\% for 6, 8, and 10 layers respectively) compared to the cases when we use the oracle utterance length (72.2\%, 73.6\%, and 69.8\% for 6, 8, and 10 layers respectively). For 10 layers, we got similar result between using the utterance length of 4000ms and the oracle utterance length with 69.8\%. This result suggests that we can use the approach even if we do not have transcription.

\section{Conclusion}

We present our study in the non-linguistic approach for detecting AD by utilizing only the speech audio data. The employment of GCNN yielded the best result of 73.6\%. Since it does not utilize linguistic information, we can easily apply it to low-resource languages.

There are still a lot of remaining tasks. In the near future, we will evaluate our current approach on different languages especially on low-resource languages. Other possible directions include estimating the severity of the disease and also evaluating its temporal change.

\section{Acknowledgements}

This work was supported by JSPS KAKENHI 16H02845 and
by JST CREST Grant Number JPMJCR1687, Japan.

\bibliographystyle{IEEEtran}



\end{document}